\documentclass[intlimits,notitlepage,9pt]{iopart}
\usepackage{graphics}
\expandafter\let\csname equation*\endcsname\relax
\expandafter\let\csname endequation*\endcsname\relax
\usepackage{amsmath,mathtools}
\usepackage{amssymb,dsfont,bbm,xfrac}
\usepackage{multirow}
\usepackage[usenames,dvipsnames]{xcolor}
\usepackage{tikz,pgfplots,bm}
\usetikzlibrary{shapes,snakes,arrows,pgfplots.groupplots,fadings,calc,decorations.markings,positioning,chains,fit,shapes,calc}
\usepackage[caption=false]{subfig}
\usepackage[english]{babel}
\usepackage{pgfplotstable}
\usepackage{hyperref}
\hypersetup{pdfauthor={Caracciolo DAchille Sicuro},pdftitle={Anomalous scaling of the optimal cost in the REAP},%
            colorlinks, linktocpage=true, pdfstartpage=3, pdfstartview=FitV,%
    breaklinks=true, pdfpagemode=UseNone, pageanchor=true, pdfpagemode=UseOutlines,%
    plainpages=false, bookmarksnumbered, bookmarksopen=true, bookmarksopenlevel=1,%
    hypertexnames=true, pdfhighlight=/O,%
    urlcolor=orange, linkcolor=blue, citecolor=red, 
        }
\usepackage[T1]{fontenc}
\usepackage{lmodern}

\DeclareMathOperator{\dd}{d}
\DeclareMathOperator{\sinc}{sinc}

\allowdisplaybreaks
\begin{document}
\title[Anomalous scaling of the optimal cost in the REAP]{Anomalous scaling of the optimal cost in the one-dimensional random assignment problem}

\author{Sergio Caracciolo}\ead{sergio.caracciolo@mi.infn.it}
\address{Dipartimento di Fisica, University of Milan and INFN, via Celoria 16, I-20133, Milan, Italy}
\author{Matteo D'Achille}\ead{matteo.d-achille@college-de-france.fr}
\address{Basic Research Division, Centre CEA de Saclay, D36 91191 Gif-sur-Yvette Cedex, France}
\address{CIRB - Center for Interdisciplinary Research in Biology, Coll\`ege de France, 11 Place Marcelin Berthelot, 75231 Paris, France}
\address{LI-PaRAD - Laboratoire d'Informatique Parall\'elisme R\'eseaux Algorithmes Distribués, 45 Avenue des \'Etats-Unis, 78035 Versailles,  France 
Universit\'e de Versailles Saint-Quentin-en-Yvelines - Universit\'e Paris Saclay}
\author{Gabriele Sicuro}\ead{gabriele.sicuro@roma1.infn.it}
\address{Dipartimento di Fisica, Sapienza Universit\`a di Roma, P.le A. Moro 2, I-00185, Rome, Italy}

\date{\today}
\begin{abstract}We consider the random Euclidean assignment problem on the line between two sets of $N$ random points, independently generated with the same probability density function $\varrho$. The cost of the matching is supposed to be dependent on a power $p>1$ of the Euclidean distance of the matched pairs. We discuss an integral expression for the average optimal cost for $N\gg 1$ that generalizes a previous result obtained for $p=2$. We also study the possible divergence of the given expression due to the vanishing of the probability density function. The provided regularization recipe allows us to recover the proper scaling law for the cost in the divergent cases, and possibly some of the involved coefficients. The possibility that the support of $\varrho$ is a disconnected interval is also analysed. We exemplify the proposed procedure and we compare our predictions with the results of numerical simulations.
\end{abstract}
\section{Introduction}
\label{intro}
Random Euclidean assignment problems (REAPs) are combinatorial optimization problems in which we couple points randomly generated on a given $d$-dimensional domain $\Omega\subseteq\mathds R^d$ in such a way that a certain cost function is minimized. To be more precise, let us call $\Xi\coloneqq\{\mathbf x_i\}_{i=1,\dots,N}\subset\Omega$ and $\Upsilon\coloneqq\{\mathbf y_j\}_{j=1,\dots,N}\subset\Omega$ two sets of points, both of cardinality $N$, in the $d$-dimensional domain $\Omega$. All points are supposed to be identically and independently distributed over $\Omega$  according to some probability density function (PDF) $\varrho\colon\Omega\to\mathds R^+$. We also suppose that a certain weight function $w\colon\Omega\times\Omega\to\mathds R^+$ is given, in such a way that $w(\mathbf x_i,\mathbf y_j)\equiv w\left(\|\mathbf x_i-\mathbf y_j\|\right)$ is the cost of the coupling $(\mathbf x_i,\mathbf y_j)$, $\|\mathbf x_i-\mathbf y_j\|$ being the Euclidean distance between $\mathbf x_i$ and $\mathbf y_j$. Then, denoting by $\mathcal S_N$ the set of permutations $\sigma$ of $N$ elements, we search for an \textit{optimal assignment} $\sigma_*$, such that
\begin{equation}\label{epsilongenerico}
 \varepsilon[\Xi,\Upsilon]\coloneqq \min_{\sigma\in\mathcal S_N}\frac{1}{N}\sum_{i=1}^N w\left(\mathbf x_{i},\mathbf y_{\sigma(i)}\right)\equiv \frac{1}{N}\sum_{i=1}^N w\left(\mathbf x_{i},\mathbf y_{\sigma_*(i)}\right).
\end{equation}
Observe that an optimal assignment always exists but it is not unique in general. From the computational point of view, once the sets of points $\Xi$ and $\Upsilon$ and the weight function $w$ are given, finding the minimum cost permutation is an ``easy'' problem, being the assignment problem in the $\mathsf P$ complexity class \cite{papadimitriou1998combinatorial}. However, due to the randomness of the point positions, we are typically interested in the \textit{average} properties of the solution, e.g., in the average optimal cost (AOC)
\begin{equation}\label{epsilongenerico2}
 \varepsilon_N\coloneqq\overline{\varepsilon[\Xi,\Upsilon]}=\overline{\min_{\sigma\in\mathcal S_N}\frac{1}{N}\sum_{i=1}^N w\left(\mathbf x_{i},\mathbf y_{\sigma(i)}\right)},
\end{equation}
where we have denoted by $\overline{\bullet}$ the average over all possible realizations of the point sets $\Xi$ and $\Upsilon$ over $\Omega$ according to the law $\varrho$.

In the available literature on the REAP, the domain $\Omega$ is typically supposed to be \textit{connected}, and the points to be \textit{uniformly distributed} over it. Many results are available for this specific formulation of the problem, in particular assuming the weight function
\begin{equation}\label{costowp}
 w_p(\mathbf x,\mathbf y)\coloneqq\|\mathbf x-\mathbf y\|^p,\quad p\in\mathds R^+.
\end{equation}
For example, in Ref.~\cite{Ajtai1984} the scaling of the AOC in the REAP with weight function $w_p$ on the unit square has been derived, whereas in Ref.~\cite{Caracciolo2014, Caracciolo2015} the coefficient to this scaling has been obtained for $p=2$ both on the unit square and on the torus, a result later rigorously proved in Ref.~\cite{Ambrosio2016}. Higher dimensions have been considered in Refs.~\cite{Caracciolo2014, Caracciolo2015}, where the weight function $w_2$ has been adopted, again assuming uniformly distributed random points on the $d$-dimensional hypercube. The problem greatly simplifies in the $d\to+\infty$ limit, to be taken before the $N\to +\infty$ limit, and indeed the infinite dimensional case has been the first to be considered in the physics literature. The simplification is due to the fact that, in this limit, the correlation among the weights $w_p(\mathbf x_i,\mathbf y_j)$ disappears, and the average over the disorder can be easily tackled using the replica trick \cite{mezard1987spin}. This fact has been first shown in a series of seminal works by Orland \cite{Orland1985} and M\'ezard and Parisi \cite{Mezard1985,Mezard1987}. In their treatment of the problem, M\'ezard and Parisi considered $p\to+\infty$ as well, in such a way that $\sfrac{d}{p}=r+1$ for some $r\geq 0$. They proved that, in this mean-field version of the original Euclidean problem, the AOC depends on $r$ only. Their exact results have been later rigorously proved \cite{Aldous1992,Linusson2004,Nair2005}, and their techniques have been extended to calculate both finite-size corrections in the mean-field model \cite{Mezard1987,Parisi2002} and finite-dimensional corrections \cite{Mezard1988,Houdayer1998,Lucibello2017}. The results obtained by M\'ezard and Parisi on random matching problems paved the way to the application of statistical physics techniques, in particular from the theory of disordered systems, to the study of combinatorial optimization problems in presence of randomness \cite{mezard1987spin}. Since their original contributions, the replica trick and the cavity method have been applied to many random optimization problems, and inspired new efficient algorithms for their solution, such as the belief propagation algorithm \cite{Bayati2008,mezard2009information}.

The REAP is also naturally related to the Monge-Kantorovich transportation problem between probability measures in the continuum \cite{Villani2009}. Indeed, it can be restated as a transportation problem between empirical measures obtained from the same PDF on an Euclidean domain \cite{bobkov2014,McCann1999}. This correspondence has been fruitfully exploited to obtain some information on the asymptotic behavior of the AOC of the REAP for $d\geq 1$ \cite{Caracciolo2014,Caracciolo2015,bobkov2014}.

Moving from these recent results, in the present paper, we study the REAP in the lowest possible dimension, i.e., $d=1$. The one-dimensional REAP is particularly simple to solve when convex weight functions $w$ are considered. Using the weight function in Eq.~\eqref{costowp} with $p>1$, and assuming the points uniformly distributed on the interval, the problem has been discussed and solved in Refs.~\cite{Boniolo2012,Caracciolo2017b,Caracciolo2014c}, where a special correspondence between the optimal permutation and a Brownian bridge process on the interval has been found in the thermodynamical limit \cite{Boniolo2012,delBarrio2005}. The $p<0$ case is discussed in Ref.~\cite{Caracciolo2017b}. For results and details on the $0<p<1$ case, see for example Refs.~\cite{McCann1999,Boniolo2012}. By means of the aforementioned relation between the REAP and the Monge-Kantorovich theory, the treatment for $p>1$ in the uniform distribution case given in Ref.~\cite{Caracciolo2014c} has been later extended to a generic distribution on the line, and a general formula for the AOC has been given for the case of the REAP, assuming a quadratic weight function $w_2$ \cite{Caracciolo2015}. A generalization of this formula, for a generic value of $p>1$, is given in Ref.~\cite{bobkov2014}. As stressed by the authors themselves in Refs.~\cite{Caracciolo2015,bobkov2014}, the provided general formula predict $\varepsilon_N=O\left(N^{-\sfrac{p}{2}}\right)$ but it may present some issues, in particular can be divergent depending on the properties of $\varrho$. In this case, the expression must be properly regularized, and a different, anomalous asymptotic behavior of $\varepsilon_N$ may appear \cite{Caracciolo2015}. Here we will present a re-derivation of such a general expression for the AOC in the REAP on the line, extending the arguments in Refs.~\cite{Caracciolo2015,Caracciolo2014c} to any value of $p>1$ and, moreover, we will give details on the delicate regularization procedure when the scaling of $\varepsilon_N$ is anomalous. 

The paper is organized as follows. In Section~\ref{sec:1} we will present the general theory for the REAP on the line. In Section~\ref{sec:2} we will discuss the problem of the anomalous scaling of the AOC in the case of gapless support of $\varrho$, we will give the recipe for the regularization procedure to be applied to the general formula presented in Section~\ref{sec:1}, and we will exemplify our results studying different types of PDFs and deriving the scaling of the AOC in all cases. In Section~\ref{sec:disconnesso} we will consider the effect of the presence of a gap in the support of $\varrho$. Finally, in Section~\ref{sec:3} we will give our conclusions.

\section{The random Euclidean assignment problem on the line}\label{sec:1}
Let us start introducing the REAP on the line and discussing its specific properties. As stated in the Introduction, we will consider a PDF $\varrho(x)\colon\mathds R\to\mathds R^+$ on the real line, $\int_{-\infty}^{+\infty}\varrho(x)\dd x=1$. Here and in the following we denote by 
\begin{equation}
 \Omega\coloneqq\{x\in\mathds R|\varrho(x)>0\}
\end{equation}
the support of $\varrho$, and therefore $\varrho(x)=0$ $\forall x\in\mathds R\setminus\Omega$. We will also denote by $\bar\Omega$ the closure of $\Omega$, possibly including the points at infinity. The cumulative function $\Phi(x)$ and the complementary cumulative $\bar\Phi(x)\coloneqq 1-\Phi(x)$ are given by
\begin{equation}
 \Phi(x)\coloneqq\int_{-\infty}^x\varrho(\xi)\dd\xi\eqqcolon1-\bar\Phi(x).
\end{equation}
Let us suppose now that two sets of points both of cardinality $N$, $\Xi\coloneqq\{x_i\}_{i=1,\dots,N}$ and $\Upsilon\coloneqq\{y_i\}_{i=1,\dots,N}$, are generated on the line, independently and with the same PDF $\varrho$. We will assume that the points in $\Xi$ and the points in $\Upsilon$ are labeled in such a way that $x_i<x_{i+1}$ and $y_{i}<y_{i+1}$ for $i=1,\dots,N-1$. We search for the optimal permutation $\sigma_*\in\mathcal S_N$ such that
\begin{equation}
 \varepsilon[\Xi,\Upsilon]\coloneqq \min_{\sigma\in\mathcal S_N}\frac{1}{N}\sum_{i=1}^N|y_{\sigma(i)}-x_i|^p\equiv \frac{1}{N}\sum_{i=1}^N|y_{\sigma_*(i)}-x_i|^p,\quad p\in\mathds R.
\end{equation}
In optimal transport theory, the optimal cost $\varepsilon[\Xi,\Upsilon]$ is the Kantorovich distance between the two empierical measures $\varrho_\Xi(x)\coloneqq\sfrac{1}{N}\sum_{i=1}^N\delta(x-x_i)$ and $\varrho_\Upsilon(x)\coloneqq\sfrac{1}{N}\sum_{i=1}^N\delta(x-y_i)$, both having $\varrho$ as (weak) limit. It turns out that, independently from the PDF $\varrho$ adopted, the optimal permutation is simply given by $\sigma_*(i)=i$, provided that $p>1$ \cite{bobkov2014,McCann1999,Boniolo2012}. This result greatly simplifies the calculation. If we introduce the transport field
\begin{equation}\label{transport}
 \phi_k\coloneqq y_k-x_k,
\end{equation}
then we can write
\begin{equation}
 \varepsilon_N\coloneqq\overline{\varepsilon[\Xi,\Upsilon]}=\frac{1}{N}\sum_{k=1}^N\int |\phi|^p \Pr[\phi_k\in\dd\phi].
\end{equation}
where we have used the notation $z\in\dd x\Leftrightarrow z\in(x,x+\dd x)$. The distribution of the $k$th transport field $\phi_k$ can be obtained observing that
\begin{equation}
 \Pr[x_k\in\dd x]=\binom{N}{k}\bar\Phi^{N-k}(x)\dd\Phi^k(x),
\end{equation}
and therefore
\begin{equation}\label{distphi}
\Pr[\phi_k\in\dd\phi]=\dd\phi \binom{N}{k}^2\iint \delta(\phi-y+x)\bar\Phi^{N-k}(x)\bar\Phi^{N-k}(y)\dd\Phi^k(x)\dd\Phi^k(y).
\end{equation}
The evaluation of $\varepsilon_N$ can be performed writing \footnote{To obtain Eq.~\eqref{costo} we have introduced the Gauss hypergeometric function 
\[\textstyle {}_2F_1[a,b;c;z]\coloneqq \sum_{k=0}^\infty\frac{(a)_k(b)_k}{(c)_k}\frac{z^k}{k!},\quad (x)_k\coloneqq \prod_{n=0}^{k-1}(x+n),\]
and we have used the fact that
\[\textstyle \sum_{k=1}^N\binom{N}{k}^2 k^2z^{k}=N^2z\sum_{k=0}^\infty\frac{(1-N)_k(1-N)_k}{(1)_k}\frac{z^k}{k!}=N^2z\,{}_2F_1\left[1-N,1-N;1;z\right].\]}
\begin{subequations}\label{costo}
\begin{align}
 \varepsilon_N=&\,\frac{1}{N}\iint_{\Omega\times\Omega} |y-x|^p \sum_{k=1}^N \binom{N}{k}^2\bar\Phi^{N-k}(x)\bar\Phi^{N-k}(y)\dd\Phi^k(x)\dd\Phi^k(y)\\
 =&\frac{1}{N}\iint_{\Omega\times\Omega} |y-x|^p\,{}_2F_1\left[1-N,1-N;1;\frac{\Phi(x)}{\bar\Phi(x)}\frac{\Phi(y)}{\bar\Phi(y)}\right]\dd\bar \Phi^N(x)\dd\bar \Phi^N(y).
\end{align}
\end{subequations}
Up to now, no approximation has been performed. Being interested in the case of very large number of points, we observe that a nontrivial large $N$ limit of Eq.~\eqref{distphi} can be obtained setting $k=Ns+\sfrac{1}{2}$ and introducing the variables $\xi$ and $\eta$ such that
\begin{subequations}\label{phixietas}
\begin{equation}\label{phixieta}
 \Phi(x)=s+\frac{\xi}{\sqrt N},\quad \Phi(y)=s+\frac{\eta}{\sqrt N},
\end{equation}
in such a way that $s$ is kept fixed when $N\to +\infty$. This rescaling has a clear interpretation if we observe that an optimal assignment configuration between $\Xi$ and $\Upsilon$ for $p>1$ can be mapped, through the cumulative function $\Phi$, to an optimal assignment configuration of the same type between points uniformly distributed on $[0,1]$, being $\Phi$ ordering preserving. As shown in Refs.~\cite{Caracciolo2017b,Caracciolo2014c}, the optimal assignment between random points uniformly distributed on the unit interval is asymptotically equivalent to a Brownian bridge process after a rescaling of the type in Eq.~\eqref{phixieta} is performed. This also implies that the (rescaled) transport field itself can be expressed, in the $N\to +\infty$ limit, in terms of the Brownian bridge process composed with the (inverse) cumulative function $\Phi^{-1}$. Assuming that $\Omega=\bar\Omega$ and that $\bar\Omega$ is connected --- i.e., that $\left(\varrho\circ\Phi^{-1}\right)(s)\neq 0$ for all $s\in[0,1]$ ---, we have
\begin{equation}
 \Phi^{-1}\left(s+\frac{\xi}{\sqrt N}\right)=\Phi^{-1}(s)+\frac{\xi}{\sqrt N \Psi_\varrho(s)}+o\left(\frac{1}{\sqrt{N}}\right),
\end{equation}
where we have introduced
\begin{equation}
 \Psi_\varrho(s)\coloneqq \left(\varrho\circ\Phi^{-1}\right)(s).
\end{equation}
A similar equation holds for $\Phi^{-1}\left(s+\sfrac{\eta}{\sqrt N}\right)$. This fact suggests us the following rescaling of $\phi_k$ to obtain a nontrivial $N\to+\infty$ limit,
\begin{equation}
 \phi_k=\frac{\varphi(s)}{\sqrt N}.
\end{equation}\end{subequations}
We remark here that
\begin{equation}\label{conds}
 \frac{1}{2N}\leq s\leq 1-\frac{1}{2N},
\end{equation}
a fact that will have important consequences in the following. We obtain, at the leading order, the following PDF for $\varphi$,
\begin{equation}\label{distphis}
\begin{split}
 \Pr[\varphi(s)\in\dd\varphi]=&\dd\varphi\iint\delta\left(\varphi-\frac{\eta-\xi}{\varrho\left(\Phi^{-1}(s)\right)}\right)\frac{\exp\left(-\frac{\xi^2+\eta^2}{2 s(1-s)}\right)}{2\pi s(1-s)}\dd\xi\dd\eta\\
=&\dd\varphi\frac{\Psi_\varrho(s)}{2\sqrt{\pi s(1-s)}}\exp\left\{-\frac{\left[\Psi_\varrho(s)\right]^2}{4 s(1-s)} \varphi ^2\right\},
\end{split}\end{equation}
that implies that
\begin{subequations}\label{costoasintotico}
\begin{align}
 N^{\sfrac{p}{2}}\varepsilon_N&=\int_0^1\dd s\frac{s^\frac{p}{2}(1-s)^\frac{p}{2}}{\left[\Psi_\varrho(s)\right]^{p}}\int_{-\infty}^{+\infty}|\varphi|^p\frac{\e^{-\frac{\varphi^2}{4}}}{2\sqrt{\pi}}\dd\varphi +o(1)\\
 &=\frac{2^p}{\sqrt\pi}\Gamma\left(\frac{p+1}{2}\right)\int_0^1\left[\frac{\sqrt{s(1-s)}}{\Psi_\varrho(s)}\right]^{p}\dd s+o(1)
 \label{costoas}\\
 &=\frac{2^p}{\sqrt\pi}\Gamma\left(\frac{p+1}{2}\right)\int_\Omega \dd x\frac{\Phi^\frac{p}{2}(x)\bar\Phi^\frac{p}{2}(x)}{\varrho^{p-1}(x)}+o(1),
\end{align}
that appears as a generalization of the expression obtained, using a different approach, in Ref.~\cite{Caracciolo2015} for $p=2$. This result can be stated in a slightly different way saying that, if $\varrho(x)$ has compact and connected support, then
\begin{equation}
 \varepsilon_N=\frac{2^p}{N^\frac{p}{2}\sqrt\pi}\Gamma\left(\frac{p+1}{2}\right)\int_\Omega \dd x\frac{\Phi^\frac{p}{2}(x)\bar\Phi^\frac{p}{2}(x)}{\varrho^{p-1}(x)}+o\left(\frac{1}{N^\frac{p}{2}}\right).\label{costoas2}
\end{equation}
\end{subequations}
The finiteness of the integral appearing in Eq.~\eqref{costoas2} guarantees the typical scaling $\varepsilon_N=O\left(N^{-\sfrac{p}{2}}\right)$ of the AOC for $N\gg 1$. We say that the AOC has an \textit{anomalous} scaling whenever the integral diverges. We will show now how information on the anomalous scaling can be extracted from the very same expression in Eqs.~\eqref{costoas2} by means of a proper regularization.

\section{The problem of regularization}\label{sec:2}
The recipe provided by Eqs.~\eqref{costoasintotico} for the calculation of the asymptotic AOC might fail due to the presence of divergences, that we have neglected assuming $\Omega=\bar\Omega$ connected. This is indeed the case for some PDFs. To explore this possibility, we will now relax the condition $\Omega=\bar\Omega$, but not the assumption that the closure $\bar\Omega$ is connected. The set $\bar\Omega\setminus\Omega$ is therefore given at most by isolated points (possibly at infinity). We will consider a disconnected $\bar\Omega$ in Section \ref{sec:disconnesso}.

The divergence of the expression in Eq.~\eqref{costoasintotico} suggests that $\lim_NN^{\sfrac{p}{2}}\varepsilon_N =+\infty$, but gives no hints about the scaling of $N^{\sfrac{p}{2}}\varepsilon_N$ with $N$. In this case, a regularization must be performed to take into account the discrete nature of the problem, i.e., the finiteness of $N$. Such a regularization will allow us to extract information on the anomalous scaling of the AOC and, possibly, on the coefficients appearing in the scaling law. Under the hypothesis $\Omega\neq\bar\Omega$ with $\bar\Omega$ connected, the expression in Eq.~\eqref{costoas} may diverge due to the presence of a point $x_*\in\partial\Omega$ (possibly at infinity) such that $\lim_{x\to x_*}\varrho(x)=0$. In particular, denoting by $s_*=\lim_{x\to x_*}\Phi(x)\in[0,1]$, a non-integrable divergence appears in Eq.~\eqref{costoas} if 
\begin{equation}\label{criterion}
 \Psi_\varrho(s)=\begin{cases}
                                O\left(s^{\sfrac{1}{2}+\sfrac{1}{p}}\right)&\text{ if $s_*=0$,}\\
                                O\left(|s-s_*|^{\sfrac{1}{p}}\right)&\text{ if $0<s_*<1$,}\\
                                O\left((1-s)^{\sfrac{1}{2}+\sfrac{1}{p}}\right)&\text{ if $s_*=1$.}
                               \end{cases}
\end{equation}
Assuming that $\bar \Omega$ is connected, $\Omega\neq\bar \Omega$ does not automatically imply the presence of an anomalous scaling of the AOC: this is therefore a necessary, but not sufficient, condition. 

The divergence can be avoided by means of a cut-off. The correct cut-off to be adopted is suggested by the approximations we have performed to obtain Eqs.~\eqref{costoasintotico} from Eq.~\eqref{costo}, that is an exact expression. 

A first regularization rule is obtained by taking into account Eq.~\eqref{conds} and therefore substituting $\int_0^1\dd s\to \int_{\sfrac{c_0}{N}}^{1-\sfrac{c_1}{N}}\dd s$ in Eq.~\eqref{costoas}, where $c_0$ and $c_1$ are two positive regularizing constants that are unspecified at this level. This regularization is required to obtain the proper leading scaling of the asymptotic AOC only if a nonintegrable singularity appears in the integral in Eq.~\eqref{costoas} at $s_*=0$ and/or $s_*=1$, and it provides information on the scaling of the $o(1)$ corrections otherwise. 

If a nonintegrable pole $s_*\in(0,1)$ is present, Eq.~\eqref{phixieta} suggests to incorporate a finite-size correction removing an open ball centered in $s_*$ having radius $\sfrac{c_*}{\sqrt N}$ for some positive regularizing constant $c_*$ to be determined. Indeed, in Eq.~\eqref{phixieta} we have approximated the quantity $\Phi(x_k)$, image of the position of the $k$th point through the cumulative $\Phi$, with its average value $s=\sfrac{k}{N}-\sfrac{1}{2N}$, introducing an error that scales as $O\left(\sfrac{1}{\sqrt{N}}\right)$. 

In all cases, it is clear that the coefficients in the AOC scaling obtained after the regularization will depend on the introduced regularizing constants, that have to be determined by means of a fit procedure. We will give now some examples of the approach described above, comparing the obtained predictions with the results of numerical simulations.

\subsection{Applications}
We apply now the considerations above to a series of PDFs $\varrho$ and we derive the asymptotic scaling of the corresponding AOC, always assuming a weight function in the form in Eq.~\eqref{costowp} with $p>1$. As preliminary observation, note that, given a REAP with PDF $\varrho$ and AOC $\varepsilon_N$, introducing the variable $x\to\hat x=\lambda x+\eta$ with $\lambda>0$ and $\eta\in\mathds R$, it is easily seen that the old problem is mapped into a new problem with $\hat \varrho(\hat x)=\frac{1}{\lambda}\varrho(\frac{\hat x-\eta}{\lambda})$ and $\hat\varepsilon_N=\lambda^p\varepsilon_N$. The results below can be therefore easily extended to more general PDFs involving nontrivial distribution parameters.

\subsubsection{Absence of singularity: flat distribution} Let us start from the simplest case, namely the REAP on the unit interval $\Omega=[0,1]$ with points extracted with uniform distribution $\varrho(x)=\theta(x)\theta(1-x)$. Here and in the following $\theta(x)$ is the Heaviside function, such that $\theta(x)=1$ if $x\geq 0$, $\theta(x)=0$ otherwise. Many results, both numerical and analytical, are available in the literature on this case \cite{Ambrosio2016,Boniolo2012,Caracciolo2017b}, both with open and with periodic boundary conditions, in particular for $p=2$ \cite{Caracciolo2014,Caracciolo2014c,Caracciolo2015a}. The flat PDF case requires no regularization, being $\Psi_\varrho(s)=1$ and therefore we will briefly recall the final result only as an application of Eq.~\eqref{costoas} for the sake of completeness and for comparison with other cases studied below. The integral in Eq.~\eqref{costoas} converges for any $p>-2$ and we obtain
\begin{equation}\label{costouniforme}
 \varepsilon_N=\frac{\Gamma\left(\sfrac{p}{2}+1\right)}{p+1}\frac{1}{N^{\sfrac{p}{2}}}+o\left(\frac{1}{N^{\sfrac{p}{2}}}\right).
\end{equation}
The expression above corresponds to the asymptotic AOC for the REAP only for $p>1$. The result for $p=2$, namely $\lim_NN\varepsilon_N=\sfrac{1}{3}$, can be derived by direct inspection observing that an exact formula is easily obtained for all values of $N$,
\begin{equation}
 \varepsilon_N=\frac{1}{N}\sum_{k=1}^N k^2\binom{N}{k}^2\iint_0^1 (y-x)^2 \left(1-x\right)^{N-k}\left(1-y\right)^{N-k}x^{k-1}y^{k-1}\dd x\dd y=\frac{1}{3}\frac{1}{N+1}.
\end{equation}
Numerical evidences for the formulas above can be found in the literature \cite{Caracciolo2014,Boniolo2012,Caracciolo2017b}, see also Fig.~\ref{fig:seno1}.

\subsubsection{Singularity for $s_*\in\{0,1\}$} Let us now consider a set of examples in which the REAP is given on the line, and such that $\Psi_\varrho(s_*)=0$ for a value $s_*\in\{0,1\}$. We consider both exponentially decaying PDFs (the exponential distribution and the Rayleigh distribution) and power-law decaying PDFs (Pareto laws), deriving the proper scaling of the AOC in all discussed cases.

\begin{figure}
\subfloat[Values of $\varepsilon_N$ in the case of exponentially distributed points, obtained for $p\in(1,2)$ compared with the theoretical prediction in Eq.~\eqref{costoasexp} (smooth line). The asymptotic value of $N^{\sfrac{p}{2}}\varepsilon_N$ for each value of $p$ has been obtained fitting the numerical results for $N$ up to $2.5\cdot 10^5$ points, assuming the scaling $f(N)=\epsilon+\epsilon_1 N^{\frac{p}{2}-1}$.\label{fig:exp1}]{\includegraphics[height=0.38\textwidth]{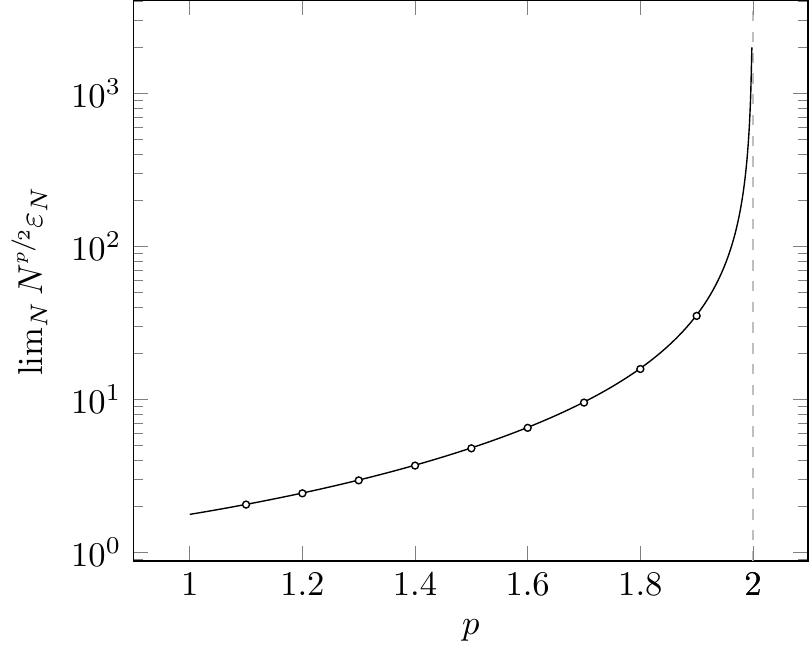}}
\hfill
\subfloat[Numerical results for the AOC for $p=2$ in the case of exponentially distributed points, showing a logarithmic divergence in agreement with the prediction in Eq.~\eqref{costoasexplog}. The smooth line is the prediction in Eq.~\eqref{costoexp2esatto}.\label{fig:exp2}]{\includegraphics[height=0.38\textwidth]{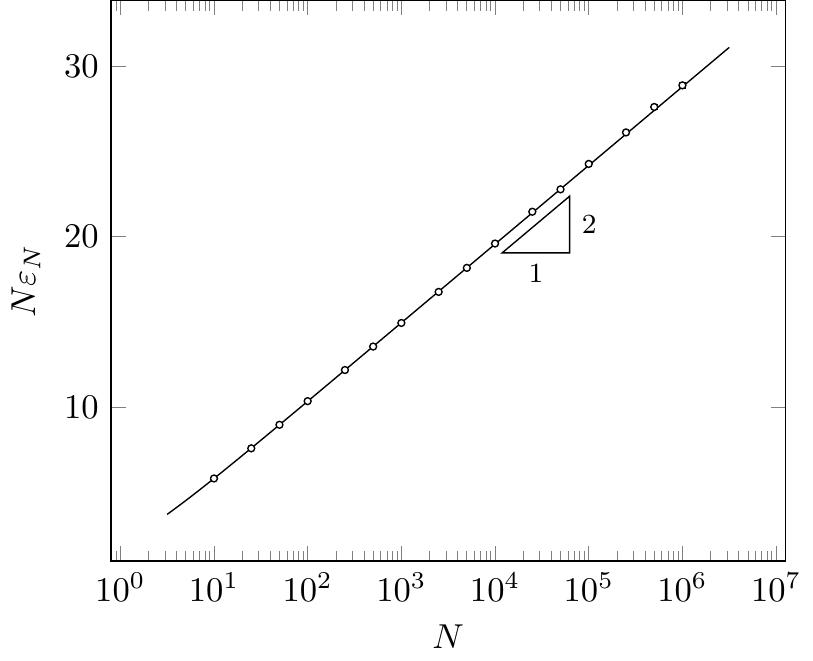}}\\
\subfloat[Numerical results for the AOC for $p>2$ in the case of exponentially distributed points. The smooth lines are fits obtained assuming the scaling behavior in Eqs.~\eqref{costoasexpg}.\label{fig:exp3}]{\includegraphics[height=0.38\textwidth]{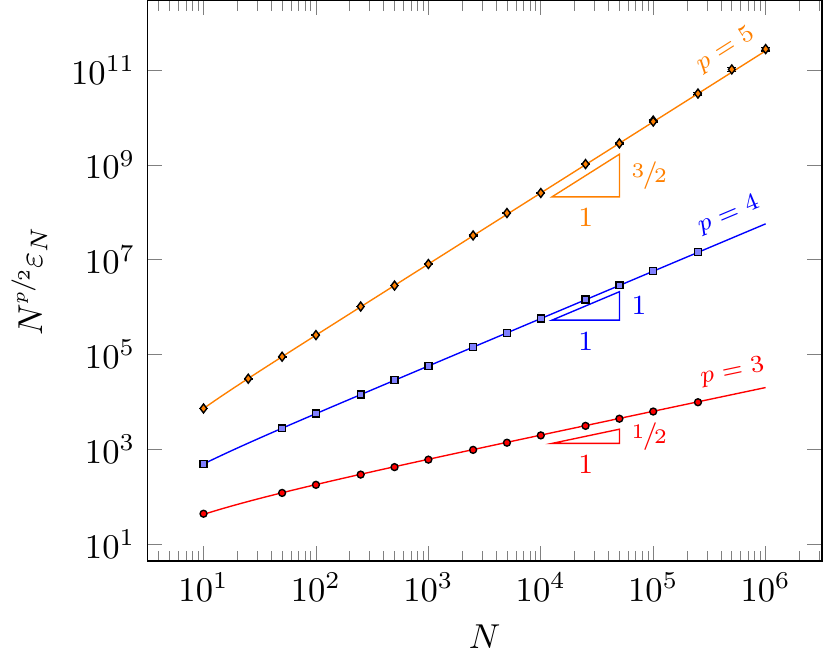}}
\hfill
\subfloat[Plot for ${\exp\left[\exp\left(N\varepsilon_N\right)\right]}\sim N^{\e^{\gamma_\text{E}}}$ in the case of Rayleigh distribution and $p=2$, to be compared with the theoretical prediction in Eq.~\eqref{costoRay}. \label{fig:Rayl}]{\includegraphics[height=0.38\textwidth]{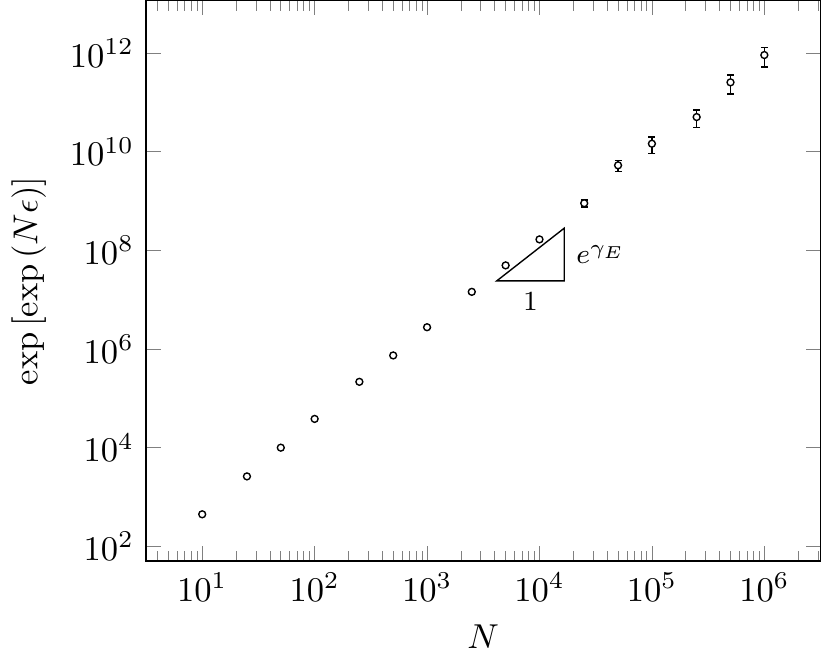}}
\caption{AOC in the case of exponential distribution and Rayleigh distribution. Error bars are represented but smaller than the markers.}
\end{figure}

\paragraph{Exponential distribution}
Let us start considering the exponential distribution,
\begin{equation}\label{expdist}
 \varrho(x)=\e^{- x}\theta(x),\quad \Phi(x)=\left(1-\e^{- x}\right)\theta(x).
\end{equation}
In this case, $\Omega=[0,+\infty)$ and a nonintegrable singularity may appear for $x\to+\infty$. For $1<p<2$ Eq.~\eqref{costoas} is convergent and the AOC is
\begin{equation}\label{costoasexp}
  N^{\sfrac{p}{2}}\varepsilon_N=\frac{2^p}{\sqrt\pi}\Gamma\left(\frac{p+1}{2}\right)\int_0^1\left(\frac{s}{1-s}\right)^\frac{p}{2}\dd s+o(1)=\Gamma(p+1)\Gamma\left(1-\frac{p}{2}\right)+o(1).
\end{equation}
The prediction above is fully consistent with numerical results, see Fig.~\ref{fig:exp1}. It is evident, however, that a divergence appears when $p=2$. Indeed, with reference to Eq.~\eqref{criterion}, we have that $\Psi_\varrho(s)=1-s=O\left((1-s)^{\sfrac{1}{2}+\sfrac{1}{p}}\right)$ if $p\geq 2$. To better understand the nature of this divergence, it is useful to calculate $\varepsilon_N$ directly, as in Eq.~\eqref{costo}, taking advantage of the fact that the computation can be performed exactly in this case for any value of $N$,
\begin{equation}\label{costoexp2esatto}
\begin{split}
 N\varepsilon_N &=\sum_{k=1}^N k^2 \binom{N}{k}^2\int_0^1\dd s\int_0^1\dd t\, \ln^2\frac{1-s}{1-t} (st)^{k-1}\left(1-s\right)^{N-k}\left(1-t\right)^{N-k}\\
 &=\sum_{k=1}^N\frac{2}{k}=2\mathrm H_N=2\ln N+2\gamma_\text{E}+\frac{1}{N}-\frac{1}{6N^2}+o\left(\frac{1}{N^2}\right),
 \end{split}
\end{equation}
where $\mathrm H_N$ is the $N$th harmonic number and $\gamma_\text{E}$ is the Euler's gamma constant. In Fig.~\ref{fig:exp2} we numerically verified the expression in Eq.~\eqref{costoexp2esatto}. The appearance of the divergence in our integral expression in Eq.~\eqref{costoas} is therefore due to an actual (logarithmic) divergence of $N\varepsilon_N$ for $N\to +\infty$. Following the criterion given in Section \ref{sec:1}, Eq.~\eqref{costoas} for $p\geq 2$ must be regularized as
\begin{equation}\label{costoexp} 
\begin{split}
N\varepsilon_N&\approx \frac{2^p}{\sqrt\pi}\Gamma\left(\frac{p+1}{2}\right)\int_0^{1-\sfrac{c}{N}}\left(\frac{s}{1-s}\right)^\frac{p}{2}\dd s\\
&=\frac{2^{p+1}}{(p+2)\sqrt\pi}\Gamma\left(\frac{p+1}{2}\right)\left(1-\frac{c}{N}\right)^{1+\sfrac{p}{2}}{}_2F_1\left[\frac{p}{2},\frac{p}{2}+1;\frac{p}{2}+2;1-\frac{c}{N}\right].
\end{split}
\end{equation}
The expression above must be interpreted as a regularization-dependent \textit{asymptotic} formula for the AOC. In particular, the large $N$ expansion will provide us the scaling properties of the optimal cost, up to some coefficients possibly depending on the regularization. For example, for $p=2$ the expression above becomes
\begin{equation}\label{costoasexplog}
   N\varepsilon_N= 2\ln N-2\log c-2+o(1),
\end{equation}
that is perfectly compatible with the exact formula in Eq.~\eqref{costoexp2esatto}, whereas the finite-size correction depends on the regularization $c$. With reference to Eq.~\eqref{costoexp2esatto}, we can also infer that $c=\e^{-\gamma_\mathrm{E}-1}$. For $p>2$ we can expand Eq.~\eqref{costoexp} as
\begin{subequations}\label{costoasexpg}
\begin{equation}\label{costoasexp2}
  N^{\sfrac{p}{2}}\varepsilon_N=\frac{2^{p+1}c^{1-\sfrac{p}{2}}}{(p-2)\sqrt\pi}\Gamma\left(\frac{p+1}{2}\right) N^{\frac{p}{2}-1}+o\left(N^{\frac{p}{2}-1}\right).
\end{equation}
In particular, for $2<p<4$ we have
\begin{equation}
 N^{\sfrac{p}{2}}\varepsilon_N=\frac{2^{p+1}c^{1-\sfrac{p}{2}}}{(p-2)\sqrt\pi}\Gamma\left(\frac{p+1}{2}\right) N^{\frac{p}{2}-1}+\Gamma\left(1-\frac{p}{2}\right)\Gamma(p+1)+o\left(1\right).
\end{equation}
We obtain therefore the scaling $\varepsilon_N=O\left(N^{-1}\right)$ for the leading term but we cannot give a prediction for the coefficient in front of it, due to its dependence on the regularization constant $c$. We have, instead, a complete analytic prediction for the finite-size correction. For $p=4$, a new logarithmic correction appears. In this case, indeed, our formula in Eq.~\eqref{costoexp} gives
\begin{equation}\label{costoasexp4}
  N^{2}\epsilon=\frac{12}{c}N-24\ln N+O\left(1\right).
\end{equation}
\end{subequations}
Once again, the coefficient of the leading term is unaccessible, despite the fact that the correct scaling is recovered, but we obtain a prediction of a logarithmic correction, with its coefficient. We do expect, but it is not obvious \textsl{a priori}, that the value of $c$ appearing in Eqs.~\eqref{costoasexpg} is the same that we have obtained for $p=2$. Performing a fit on our numerical results, presented in Fig.~\ref{fig:exp3}, we get $c=0.203(2)$ for $p=3$, $0.2084(4)$ for $p=4$ and $c=0.2069(5)$ for $p=5$, that are all very close to the value of $c=0.206549\dots$ analytically obtained for $p=2$.

\paragraph{Rayleigh distribution} As further example of regularization in the case of a PDF with exponential tail, we consider now the Rayleigh distribution,
\begin{equation}
 \varrho(x)=x \e^{-\frac{x^2}{2}}\theta(x),\quad\Phi(x)=\left(1-\e^{-\frac{x^2}{2}}\right)\theta(x).
\end{equation}
In this case $\Omega=(0,+\infty)$ and we have that $\Psi_\varrho(s)=(1-s)\sqrt{-2\ln(1-s)}$ that is infinitesimal both in $s=0$ and in $s=1$. In particular, $\Psi_\varrho(s)=\sqrt{2s}+O\left(s^{\sfrac{3}{2}}\right)$ for $s\to 0$ and therefore, according to Eq.~\eqref{criterion}, there are no integrability issues for $s\to 0$ for any value of $p>1$. On the contrary, for $s\to 1$ $\Psi_\varrho(s)=O\left((1-s)^{\sfrac{1}{2}+\sfrac{1}{p}}\right)$ for any $p>1$. The integral is therefore always divergent and a regularization is needed. We proceed in the usual way, restricting ourselves to the $p=2$ case,
\begin{equation}\label{costoRay}
 N\varepsilon_N\approx \int_0^{1-\frac{c}{N}}\frac{s}{s-1}\frac{1}{\ln(1-s)}\dd s
 =\gamma_\text{E}+\ln\ln\frac{N}{c}+\int_{\sfrac{N}{c}}^{+\infty}\frac{\dd z}{z^2\ln z}=\ln\ln N+\gamma_\mathrm{E}+ o(1).
\end{equation}
Our results have been numerically verified, as shown in Fig.~\ref{fig:Rayl}.

\begin{figure}
\subfloat[Values of $\varepsilon_N$ obtained for $p=1.1$ and $p=1.2$ and different values of $\alpha$, compared with the theoretical prediction in Eq.~\eqref{costoasPareto1} (smooth lines). The asymptotic value of $N^{\sfrac{p}{2}}\varepsilon_N$ for each value of $p$ has been obtained fitting the numerical results for $N$ up to $10^5$ points, assuming the scaling $f(N)=\epsilon+\epsilon_1 N^{p\frac{\alpha+2}{2\alpha}-1}$.\label{fig:Pareto1}]{\includegraphics[height=0.38\textwidth]{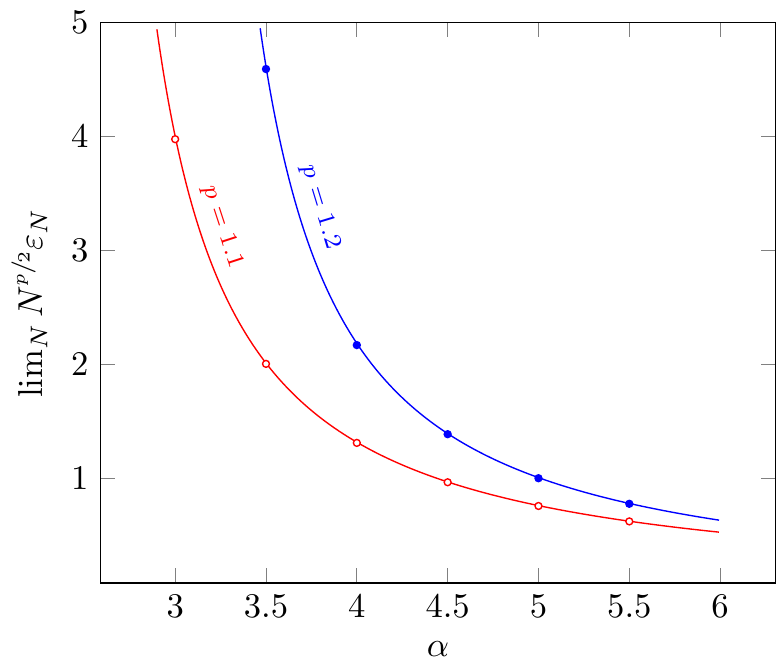}}
\hfill
\subfloat[Numerical results for the AOC for $p=2$ and different values of $\alpha$. The fits are obtained using a fitting function in the form given by Eq.~\eqref{costoasPareto2}; we obtained $c=0.0668(5)$ for $\alpha=3$, $c=0.0939(5)$ for $\alpha=4$ and $c=0.1121(6)$ for $\alpha=5$. \label{fig:Pareto2}]{\includegraphics[height=0.38\textwidth]{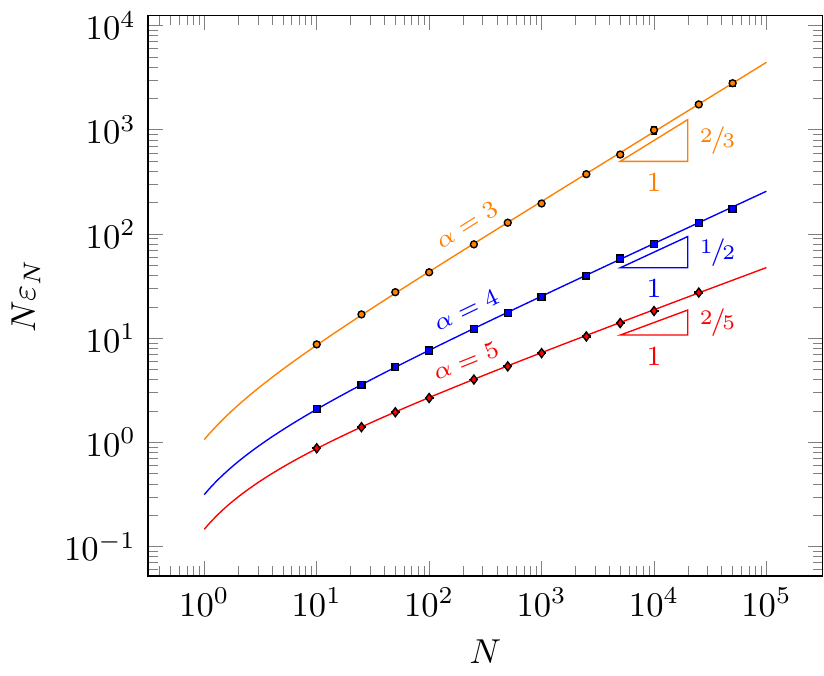}}
\caption{AOC in the case of Pareto distribution. Error bars are represented but smaller than the markers.}
\end{figure}

\paragraph{Pareto distribution} Let us now consider a power-law decaying PDF, e.g., a Pareto distribution,
\begin{equation}
 \varrho_\alpha(x)=\frac{\alpha}{x^{\alpha+1}}\theta(x-1),\quad\Phi_\alpha(x)=\frac{x^\alpha-1}{x^\alpha}\theta(x-1),\quad \alpha>0.
\end{equation}
Here we have $\Omega=[1,+\infty)$. If we consider
\begin{equation}
 p<\frac{2\alpha}{\alpha+2},
\end{equation}
Eq.~\eqref{costoas} gives a finite result, namely
\begin{equation}\label{costoasPareto1}
\begin{split}
 \lim_N N^{\sfrac{p}{2}}\varepsilon_N&=\frac{2^p}{\alpha^p\sqrt\pi}\Gamma\left(\frac{p+1}{2}\right)\int_0^1s^\frac{p}{2}(1-s)^{\frac{p}{2}-\frac{p}{\alpha}-p}\dd s\\
 &=\frac{2^p}{\alpha^p\sqrt\pi}\frac{\Gamma\left(\frac{p}{2}+1\right)\Gamma\left(1-\frac{2+\alpha}{2\alpha}p\right)\Gamma\left(\frac{p+1}{2}\right)}{\Gamma\left(2-\frac{p}{\alpha}\right)}.
 \end{split}
\end{equation}
This formula has been verified, for a set of values of $p$ and $\alpha$, in Fig.~\ref{fig:Pareto1}. When $(2-p)\alpha\leq 2p$ the integral does not converge (in particular, does not converge for any value of $\alpha$ when $p=2$). Indeed, with reference to Eq.~\eqref{criterion}, $\Psi_{\varrho_\alpha}(s)=\alpha(1-s)^{1+\frac{1}{\alpha}}$, and, therefore, a non-integrable singularity appears for $s_*=1$ when $1+\sfrac{1}{\alpha}\geq\sfrac{1}{2}+\sfrac{1}{p}$. We can proceed regularizing the integral for $p\geq \frac{2\alpha}{\alpha+2}$,
\begin{equation}
\begin{split}
N^{\sfrac{p}{2}}\varepsilon_N&\approx\frac{2^p}{\alpha^p\sqrt\pi}\Gamma\left(\frac{p+1}{2}\right)\int_0^{1-\frac{c}{N}}s^\frac{p}{2}(1-s)^{\frac{p}{2}-\frac{p}{\alpha}-p}\dd s\\
 &=\frac{2^{p+1}}{\alpha^{p}\sqrt\pi}\frac{\Gamma\left(\frac{p+1}{2}\right)}{p+2}\left(1-\frac{c}{N}\right)^{\frac{p}{2}+1}{}_2F_1\left[\frac{p}{2}+1,\frac{p}{2}\frac{\alpha+2}{\alpha};\frac{p}{2}+2;1-\frac{c}{N}\right]\\
 &=\begin{cases}
   \frac{2^{p+1}}{\alpha^{p-1}\sqrt\pi}\frac{\Gamma\left(\frac{p+1}{2}\right)}{2p+\alpha p-2\alpha}\left(\frac{N}{c}\right)^{p\frac{\alpha+2}{2\alpha}-1}+o\left(N^{p\frac{\alpha+2}{2\alpha}-1}\right)
   & \text{for $p>\frac{2\alpha}{\alpha+2}$},\\
   \frac{p+2}{2}\ln N-\frac{p+2}{2}\left(\mathrm H_{\sfrac{p}{2}}+\ln c\right)+ o(1)& \text{for $p=\frac{2\alpha}{\alpha+2}$}.
  \end{cases}
\end{split}
\end{equation}
For $p=2$, for example, we find
\begin{equation}\label{costoasPareto2}
 N\varepsilon_N=\frac{1}{\alpha}\left(\frac{N}{c}\right)^{\sfrac{2}{\alpha}}+
 \begin{cases}
  \frac{2}{\alpha(\alpha-2)} \left(\frac{N}{c}\right)^{\frac{2}{\alpha}-1}+o(N^{\frac{2}{\alpha}-1})&\text{ for $0<\alpha< 2$},\\
  -\frac{\ln N}{2}+\frac{\ln c-1}{2}+o(1)&\text{ for $\alpha=2$},\\
  -\frac{1}{\alpha-2}+o(1)&\text{ for $\alpha> 2$}.
 \end{cases}
\end{equation}
In Fig.~\ref{fig:Pareto2} we numerically verify all the scaling properties above for $p=2$ and $\alpha=3,4,5$.

\subsubsection{Singularity for $s_*\in(0,1)$} 
Let us now consider a PDF such that $\Psi_\varrho(s_*)=0$ for $0<s_*<1$ and let us derive the scaling properties of the corresponding AOC in this case. As an example, we consider
\begin{subequations}\label{cos2alpha}
\begin{align}
 \varrho_\alpha(x)&=\frac{2\cos^2(\alpha\pi x)}{1+\sinc(2\pi\alpha)}\, \theta(x)\theta(1-x),\quad\alpha\in(0,1],\\
 \Phi_\alpha(x)&=x\frac{1+\sinc(2\pi x\alpha)}{1+\sinc(2\pi\alpha)}\theta(x)\theta(1-x).
\end{align}
\end{subequations}
Here we have used the definition $\sinc(x)\coloneqq\frac{\sin(x)}{x}$. The distribution above recovers the uniform one for $\alpha\to 0$ and it has a (double) zero for $x=\sfrac{1}{2\alpha}\in[\sfrac{1}{2},1]$ if $\alpha\in[\sfrac{1}{2},1]$. The support is therefore $\Omega=[0,1]\setminus\{\sfrac{1}{2\alpha}\}$. In particular, for $\alpha\in[\sfrac{1}{2},1]$, we have
\begin{multline}
\Psi_{\varrho_\alpha}(s)=\frac{(6\pi\alpha)^{\sfrac{2}{3}}}{\sqrt[3]{2+2\sinc(2\pi\alpha)}}\left(s-s_*\right)^{\sfrac{2}{3}}+o\left(\left(s-s_*\right)^{\sfrac{2}{3}}\right),\\
\text{with }s_*\coloneqq \frac{1}{2\alpha}\frac{1}{1+\sinc(2\pi\alpha)}.
\end{multline}
Depending on the value of $\alpha$, the general exposition given in the present Section implies therefore three different regimes of the asymptotic AOC. 

\begin{figure}
\subfloat[Values of $\varepsilon_N$ in the case of points distributed with PDF given in Eq.~\eqref{cos2alpha}, obtained for different values of $p$ and $\alpha$ for which $\lim_N N^{\sfrac{p}{2}}\varepsilon_N$ is finite, compared with the theoretical prediction obtained using Eq.~\eqref{costoas} (smooth lines). The limit curve for $\alpha=0$, given by Eq.~\eqref{costouniforme}, is also represented (gray), along with the numerical results for the asymptotic AOC in the uniform distribution case. \label{fig:seno1}]{\includegraphics[height=0.38\textwidth]{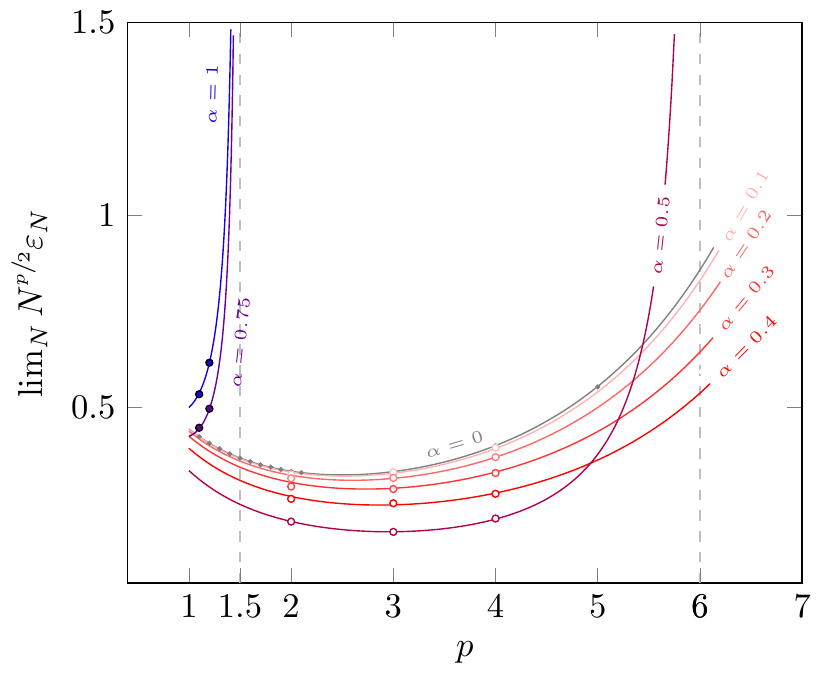}}
\hfill
\subfloat[Numerical results for the AOC in the case of points distributed with PDF given in Eq.~\eqref{cos2alpha} with $\alpha=1$. The smooth lines are fits obtained a scaling of the type $f(N)=N^{\sfrac{p}{3}-\sfrac{1}{2}}\left(\epsilon_1+\sfrac{\epsilon_2}{N}\right)+\epsilon_0$. The obtained scaling are in agreement with the prediction in Eq.~\eqref{cos10}.\label{fig:seno2}]{\includegraphics[height=0.38\textwidth]{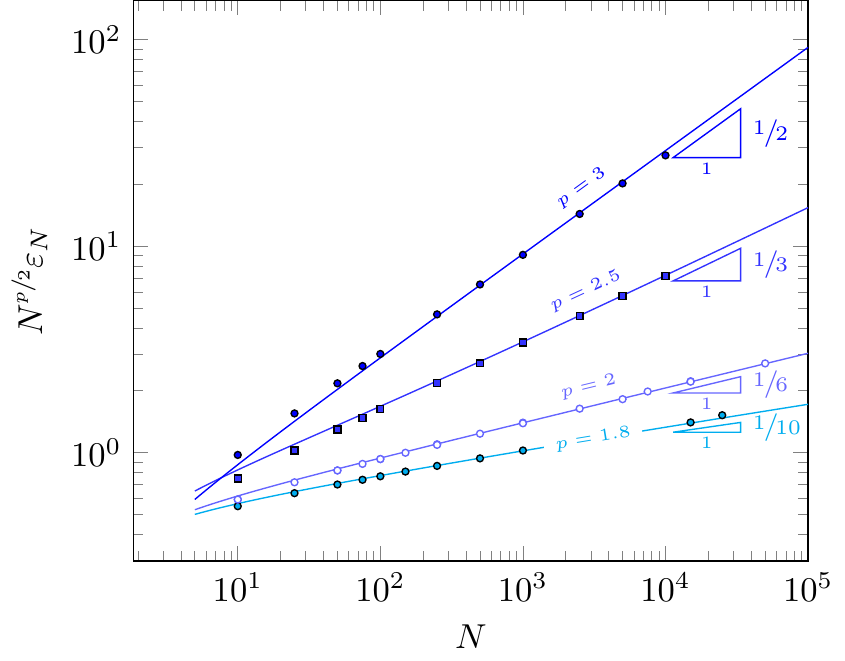}}\\
\subfloat[Numerical results for the AOC in the case of points distributed with PDF given in Eq.~\eqref{cos2alpha} with $\alpha=\sfrac{1}{2}$. The represented fits have been obtained assuming a fitting function $f(N)=N^{\sfrac{p}{6}-1}\left(\epsilon_1+\sfrac{\epsilon_2}{N}\right)+\epsilon_0$. The obtained scaling laws are in agreement with the prediction in Eq.~\eqref{cos05}.\label{fig:seno3}]{\includegraphics[height=0.38\textwidth]{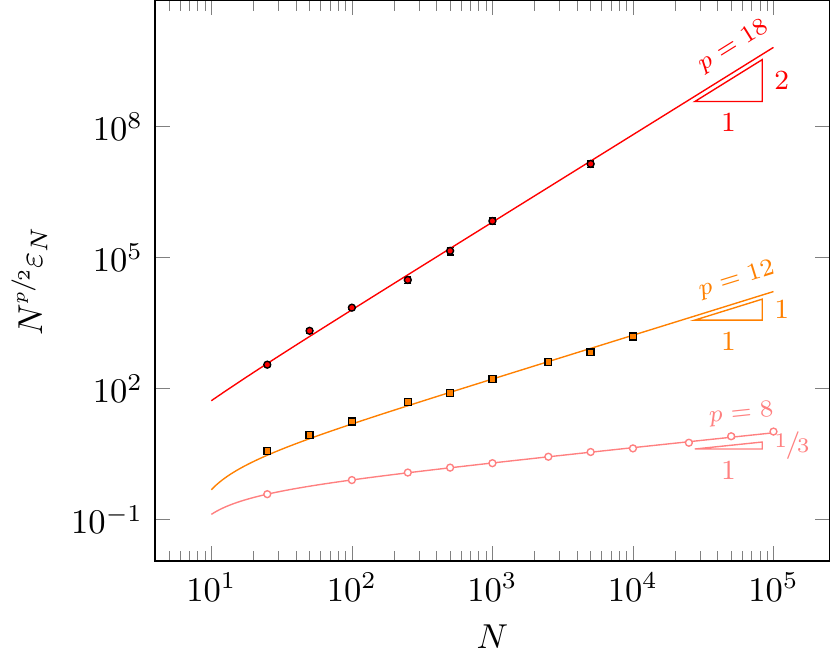}}
\hfill
\subfloat[Numerical results for the AOC in the case of points distributed with PDF given in Eq.~\eqref{cos2alpha} in the cases in which a logarithmic divergence appears. The smooth lines are fits obtained assuming a scaling $f(N)=\epsilon \ln N+\epsilon_0+\sfrac{\epsilon_1}{\ln N}$, where $\epsilon$ has been provided by the predictions in Eq.~\eqref{cos10log} and Eq.~\eqref{cos05log}. \label{fig:seno4}]{\includegraphics[height=0.38\textwidth]{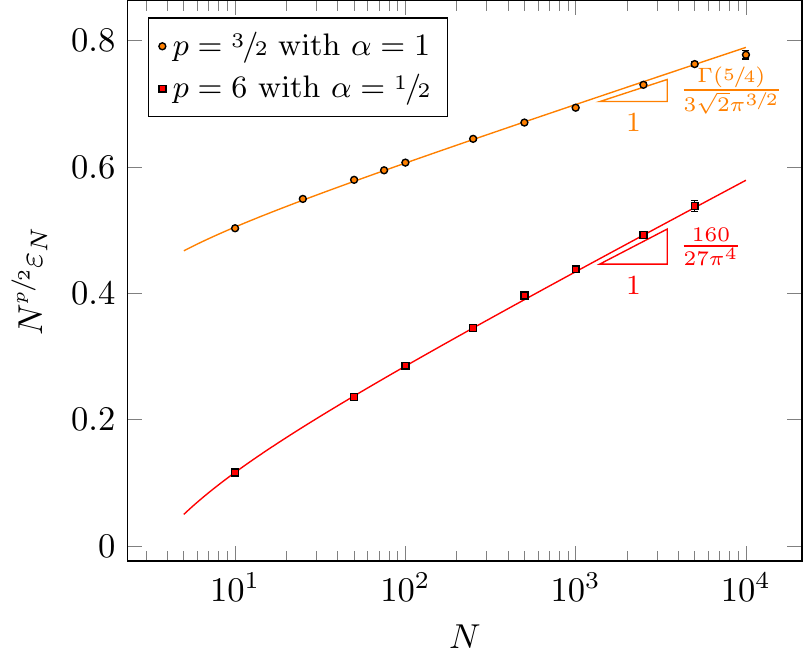}}
\caption{AOC in the case of of points distributed with PDF given in Eq.~\eqref{cos2alpha}. Error bars are represented but smaller than the markers.}
\end{figure}

For $\alpha\in(0,\sfrac{1}{2})$ the asymptotic AOC is finite for any value of $p>1$. The integral in Eq.~\eqref{costoas} has been evaluated numerically and the prediction has been compared with our numerical results in Fig.~\ref{fig:seno1}, where a perfect agreement is found.

When $\alpha=\sfrac{1}{2}$ the singularity $s_*$ moves to $1$. We obtain the regularized integral
\begin{equation}\label{cos05}
 N^{\sfrac{p}{2}}\varepsilon_N
 =\frac{2^{p}}{\sqrt\pi}\Gamma\left(\frac{p+1}{2}\right)\int_0^{1-\frac{c_{\sfrac{1}{2}}}{\sqrt[3]{N}}}\dd x\frac{\Phi_\alpha^\frac{p}{2}(x)\bar\Phi_\alpha^\frac{p}{2}(x)}{\varrho_\alpha^{p-1}(x)}+o(1)\propto \begin{cases}N^{\frac{p}{6}-1}&\text{for $p>6$,}\\
 \ln N&\text{for $p=6$,}\\
 \text{constant}&\text{for $1<p<6$.}
 \end{cases}
\end{equation}
We verified the scaling above in Fig.~\ref{fig:seno3}. In the $p=6$ case, in particular, we have
\begin{equation}\label{cos05log}
 N^{3}\varepsilon_N=\frac{160}{27\pi^4}\ln N+O(1).
\end{equation}

For $\alpha\in(\sfrac{1}{2},1]$, instead, there is a singularity in $s_*\in[\sfrac{1}{2},1)$ and the regularization procedure has to be modified. In this case we have to exclude from the integration domain a ball centered in $s_*$ and radius $O(\sfrac{1}{\sqrt N})$. Observing that
\begin{multline}
 \Phi_\alpha^{-1}\left(\frac{1}{2\alpha(1+\sinc(2\alpha\pi))}\pm\frac{c}{\sqrt N}\right)=\frac{1}{2\alpha}\pm\frac{1}{\alpha}\sqrt[3]{\frac{3c(1+\sinc(2\pi\alpha))}{2\pi^2 \sqrt N}}+o\left(\frac{1}{\sqrt[6]{N}}\right),\\
 \equiv \frac{1}{2\alpha}\pm\frac{\hat c_\alpha}{\sqrt[6]{N}}+o\left(\frac{1}{\sqrt[6]{N}}\right)
\end{multline}
and denoting the ``regularized'' domain by 
\begin{equation}
 \Omega_\alpha\coloneqq [0,1]\setminus\left(\frac{1}{2\alpha}-\frac{\hat c_\alpha}{\sqrt[6]{N}},\frac{1}{2\alpha}+\frac{\hat c_\alpha}{\sqrt[6]{N}}\right),
\end{equation}
we can write the regularized integral as
\begin{equation}\label{cos10}
 N^{\sfrac{p}{2}}\varepsilon_N
 =\frac{2^{p}}{\sqrt\pi}\Gamma\left(\frac{p+1}{2}\right)\int_{\Omega_\alpha}\dd x\frac{\Phi_\alpha^\frac{p}{2}(x)\bar\Phi_\alpha^\frac{p}{2}(x)}{\varrho_\alpha^{p-1}(x)}+o(1)\propto \begin{cases}N^{\frac{2p-3}{6}}&\text{for $p>\sfrac{3}{2}$,}\\
 \ln N&\text{for $p=\sfrac{3}{2}$,}\\
 \text{constant}&\text{for $1<p<\sfrac{3}{2}$,}
 \end{cases}
\end{equation}
where we limited ourselves to the scaling behavior of the AOC. In Fig.~\ref{fig:seno2} we present our numerical results for $p=2$ and $p=3$. In both cases, the scaling predicted in Eq.~\eqref{cos10} is confirmed. For $p=\sfrac{3}{2}$, in particular, we find
\begin{equation}\label{cos10log}
 N^{\sfrac{3}{4}}\varepsilon_N=\frac{\Gamma\left(\sfrac{1}{4}\right)}{6\alpha\left(1+\sinc(2\pi\alpha)\right)}\left(\frac{1+\sinc(2\pi\alpha)-\frac{1}{2\alpha}}{2\alpha\pi^2}\right)^{\sfrac{3}{4}}\ln N+O(1).
\end{equation}
In Fig.~\ref{fig:seno4} we show our numerical results for this case, once again in agreement with the prediction. Observe that the different regularization applied in this case implies a completely different scaling of the asymptotic AOC respect to the one obtained for $s_*=1$. 


\section{Assignment on disjoint intervals: an example}\label{sec:disconnesso}
In the examples above, and in the general remarks in Section~\ref{sec:2}, we have always assumed that the domain $\Omega$ is such that $\bar\Omega$ is a connected interval, and therefore $\Phi$ is an invertible function on $\bar\Omega$. This is not the case if the domain $\Omega$ has a ``gap''. In this Section we will study the effects of such a gap on the asymptotic AOC. We will limit ourselves to the case $\Omega=A\cup B$ with $\bar A$, $\bar B$ connected intervals such that $\bar A\cap\bar B=\emptyset$. In the following we will assume that $\forall x\in A$ and $\forall y\in B$, $x<y$. To avoid complications due to the presence of singularities in the integrals, we will also assume that $\Omega=\bar\Omega$. The lack of invertibility of $\Phi$ is due in this case to the fact that $\lim_{x\to\sup A}\Phi(x)=\lim_{x\to\inf B}\Phi(x)=\hat s$, despite the fact that $a=\sup A\neq\inf B=b$. We expect that our approach proposed in Section~\ref{sec:1} fails in this situation, because the transport field $\phi_k$ in Eq.~\eqref{transport} is not infinitesimal in general for $N\to +\infty$. 

In the simple case mentioned here, $\Omega=A\cup B$ with $\bar A\cap\bar B=\emptyset$, the exact formula in Eq.~\eqref{costo} can be written as
\begin{equation}\label{disg}
 \varepsilon_N=\frac{1}{N}\int\dd\phi\,|\phi|^p \sum_{k=1}^\infty \left[p_k^{(AA)}(\phi)+p_k^{(BB)}(\phi)+p_k^{(AB)}(\phi)+p_k^{(BA)}(\phi)\right].
\end{equation}
In the expression above, the quantity
\begin{multline}
 p_k^{(XY)}(\phi)\dd\phi\coloneqq \Pr[\phi_k\in\dd\phi,x_k\in X,y_k\in Y]\\
 =\dd\phi \binom{N}{k}^2\iint_{X\times Y}\delta(\phi-y+x)\bar\Phi^{N-k}(x)\bar\Phi^{N-k}(y)\dd\bar \Phi^k(x)\dd\bar \Phi^k(y)
\end{multline}
is the joint probability that the $k$th transport field $\phi_k=y_k-x_k$ takes value in the interval $(\phi,\phi+\dd\phi)$, $x_k\in X$ and $y_k\in Y$. We expect that, to obtain a nontrivial $N\to+\infty$ limit from $p_k^{(AA)}$ and $p_k^{(BB)}$, we have to rescale $\phi_k$ following Eq.~\eqref{phixietas}, due to the fact that matched points in the same interval can be arbitrarily close in the thermodynamical limit. Indeed, we can repeat the same calculations in Section~\ref{sec:1} performing the rescaling in Eqs.~\eqref{phixietas} and recovering, with the same \textsl{caveat}, a limiting distribution exactly in the form given in Eq.~\eqref{distphis},
\begin{multline}\label{distphi2}
 \frac{1}{N}\sum_{k=1}^{N}\Pr[\phi_k\in\dd\phi,x_k\in A,y_k\in A]+\frac{1}{N}\sum_{k=1}^{N}\Pr[\phi_k\in\dd\phi,x_k\in A,y_k\in A]\\
 =\dd\varphi\int_{0}^{1}\dd s \frac{\Psi_\varrho(s)}{2\sqrt{\pi s(1-s)}}\exp\left\{-\frac{\left[\Psi_\varrho(s)\right]^2}{4 s(1-s)} \varphi ^2\right\}+o(1).
\end{multline}
This formula is exactly the expression we would have obtained if $\bar\Omega$ were connected. If convergent, as it will happen under the hypotheses adopted here, this contribution will give a $O(N^{-\sfrac{p}{2}})$ term in the expression of $\varepsilon_N$ for $N\gg 1$.

On the other hand, the last two contributions in Eq.~\eqref{disg} corresponds to the matching transport between the two components of $\Omega$, i.e., $A\to B$ or $B\to A$, and therefore the transport field in this case is of the order of the distance between $A$ and $B$, namely $\inf B-\sup A$. The asymptotic rescaling given in Eqs.~\eqref{phixietas}, therefore, cannot be applied to this term. However, from the fact that two matched points $x_k$ and $y_k$ have $\left|\Phi(x_k)-\Phi(y_k)\right|=O\left(\sfrac{1}{\sqrt N}\right)$, if $x_k\in A$ and $y_k\in B$ this implies that
\begin{equation}
 \Phi(x_k)=\hat s+\frac{\xi_k}{\sqrt N},\quad \Phi(y_k)=\hat s+\frac{\eta_k}{\sqrt N}
\end{equation}
with $\xi_k<0$ and $\eta_k>0$, and therefore
\begin{equation}
 x_k=a+\frac{\xi_k}{\sqrt N \varrho(a)}+o\left(\frac{1}{\sqrt N}\right),\quad y_k=b+\frac{\eta_k}{\sqrt N\varrho(b)}+o\left(\frac{1}{\sqrt N}\right),
\end{equation}
where $a=\sup A$ and $b=\inf B$ and, under our hypotheses, $\varrho(a)\neq 0$ and $\varrho(b)\neq 0$. The relations above suggest the rescaling $\phi_{k}\to b-a+\sfrac{\hat \varphi_k}{\sqrt N}$ for $k=Ns+\sfrac{1}{2}$. A nontrivial distribution for $\hat\varphi$ is obtained assuming $s=\hat s+\sfrac{\sigma}{\sqrt N}$. Indeed
\begin{multline}
 \frac{1}{N}\sum_{k=1}^{N}\Pr[\phi_k\in\dd\phi,x_k\in A,y_k\in B]\\
 =\frac{\dd\phi}{N}\sum_{k=1}^N\binom{N}{k}^2\int_{0}^{\hat s}\dd u^k\int_{\hat s}^1\dd v^k\,\delta\left(\phi-\Phi^{-1}(v)+\Phi^{-1}(u)\right)(1-u)^{N-k}(1-v)^{N-k}\\
 \approx\dd\hat\varphi\int_{-\infty}^{+\infty}\frac{\dd\sigma}{\sqrt N} \int^0_{-\infty}\dd\xi \int_0^{+\infty}\dd\eta\,\delta\left(\hat\varphi-\frac{\eta}{\varrho(b)}+\frac{\xi}{\varrho(a)}\right)\frac{\exp\left(-\frac{(\xi-\sigma)^2+(\eta-\sigma)^2}{2\hat s(1-\hat s)}\right)}{2\pi \hat s(1-\hat s)}\\
 =\frac{\dd\hat\varphi}{\sqrt N}\begin{cases}
   \frac{\varrho(a)\varrho(b)}{2\left(\varrho(a)-\varrho(b)\right)}\left[\mathrm{Erf}\left(\frac{\varrho(a)\hat\varphi}{\sqrt{2\hat s(1-\hat s)}}\right)-\mathrm{Erf}\left(\frac{\varrho(b)\hat\varphi}{\sqrt{2\hat s(1-\hat s)}}\right)\right]\theta(\hat\varphi)&\varrho(a)\neq\varrho(b),\\
   \varrho^2(a)\hat\varphi\frac{\e^{-\frac{\varrho^2(a)\hat\varphi^2}{4\hat s(1-\hat s)}}}{2\sqrt{\pi \hat s(1-\hat s)}}\theta(\hat\varphi)&\varrho(a)=\varrho(b),
  \end{cases}\\
  \coloneqq \frac{1}{\sqrt N}\Pr[\hat\varphi\in\dd\hat \varphi, A\to B].
\end{multline}
In a similar manner the expression for $\Pr[\hat\varphi\in\dd\hat\varphi, B\to A]$ can be obtained. Collecting our results, we can write down the contribution to the asymptotic probability for $\hat\varphi$ given by the matching between points of different subintervals as
\begin{equation}
 \Pr[\hat\varphi(s)\in\dd\hat\varphi,A\leftrightarrow B]=\frac{\dd\hat\varphi}{\sqrt N}\begin{cases}
   \varrho(a)\varrho(b)\frac{\mathrm{Erf}\left(\frac{\varrho(b)|\hat\varphi|}{\sqrt{2\hat s(1-\hat s)}}\right)-\mathrm{Erf}\left(\frac{\varrho(a)|\hat\varphi|}{\sqrt{2\hat s(1-\hat s)}}\right)}{2\left(\varrho(b)-\varrho(a)\right)}&\varrho(a)\neq\varrho(b),\\
   \varrho^2(a)|\hat\varphi|\frac{\e^{-\frac{\varrho^2(a)\hat\varphi^2}{4\hat s(1-\hat s)}}}{2\sqrt{\pi \hat s(1-\hat s)}}&\varrho(a)=\varrho(b).
  \end{cases}
\end{equation}
Observe that the previous contributions are not normalized in $\hat\varphi$. This is due to the fact that they appears as $O(\sfrac{1}{\sqrt{N}})$ corrections to the distribution $\Pr[\varphi(s)\in\dd \varphi]$ that has Eq.~\eqref{distphi2} as leading term: higher order corrections to $\Pr[\phi_k\in\dd\phi,x_k\in A,y_k\in A]$ and $\Pr[\phi_k\in\dd\phi,x_k\in B,y_k\in B]$, that would guarantee for $N\gg 1$  the total integral of the corrections to $\Pr[\varphi(s)\in\dd \varphi]$ to be zero, have not been computed. This will be irrelevant for our final computation, because the matching field is $O(\sfrac{1}{\sqrt N})$ when matching points in the same interval, but $O(1)$ when matching points in different intervals. The final result is
\begin{multline}\label{asygap}
 \varepsilon_N=\frac{|b-a|^p}{\sqrt N}\int_{-\infty}^{+\infty}\Pr[\hat\varphi(s)\in\dd\hat\varphi,A\leftrightarrow B]+o\left(\frac{1}{\sqrt{N}}\right)\\
 =\frac{2|b-a|^p}{\sqrt N}\sqrt{\frac{\hat s(1-\hat s)}{\pi}}+o\left(\frac{1}{\sqrt{N}}\right),
\end{multline}
irrespectively from the fact that $\varrho(a)=\varrho(b)$ or not. Remarkably, the coefficient in front of the leading term does not depend on $\varrho(x)$ but only on the average fraction of points that are in each of the two subintervals, i.e., on $\hat s$. Moreover, the obtained scaling can be intuitively justified observing that the number of points of $\Xi$ that are expected to fall, e.g., in $A$ are $N\hat s$, but the fluctuations to this number scale as $\sqrt{N}$. The same reasoning applies to $\Upsilon$. This means that $O(\sqrt N)$ points in $A$ have necessarily to be matched with points in $B$ with a matching cost that is $O(|b-a|^p)$, giving a final $O(\sfrac{1}{\sqrt N})$ contribution to $\varepsilon_N$.

\begin{figure}
\subfloat[Numerical results for the AOC in the case of points distributed with PDF given in Eq.~\eqref{rhogap} with $q=\sfrac{1}{2}$ and $p=2$. The fits are obtained assuming a scaling $f(N)=\alpha^2\sqrt{\sfrac{N}{\pi}}+\epsilon$, where $\epsilon$ is a fitting parameter. The dashed line represents the asymptotic limit for $q=0$ predicted by Eq.~\eqref{costoasintotico}. \label{fig:gap1}]{\includegraphics[height=0.38\textwidth]{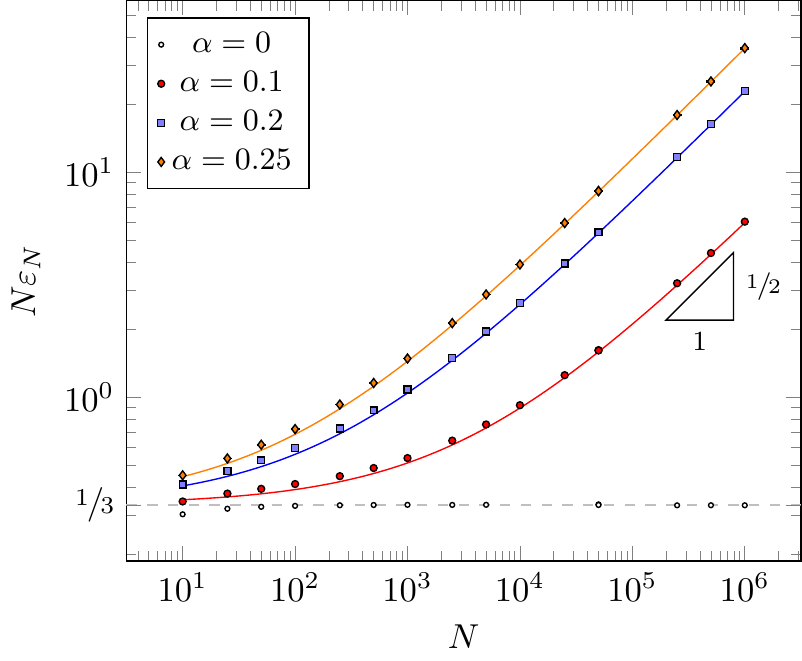}}
\hfill
\subfloat[Numerical results for the AOC in the case of points distributed with PDF given in Eq.~\eqref{rhogap} with $q=\sfrac{3}{4}$ and and $p=4$. The fits are obtained assuming a scaling $f(N)=\sfrac{1}{4}\sqrt{\sfrac{3}{\pi}}\alpha^4N^{\sfrac{3}{2}}+\epsilon\sqrt N$, where $\epsilon$ is a fitting parameter \label{fig:gap2}. The dashed line represents the asymptotic limit for $q=0$ predicted by Eq.~\eqref{costoasintotico}.]{\includegraphics[height=0.38\textwidth]{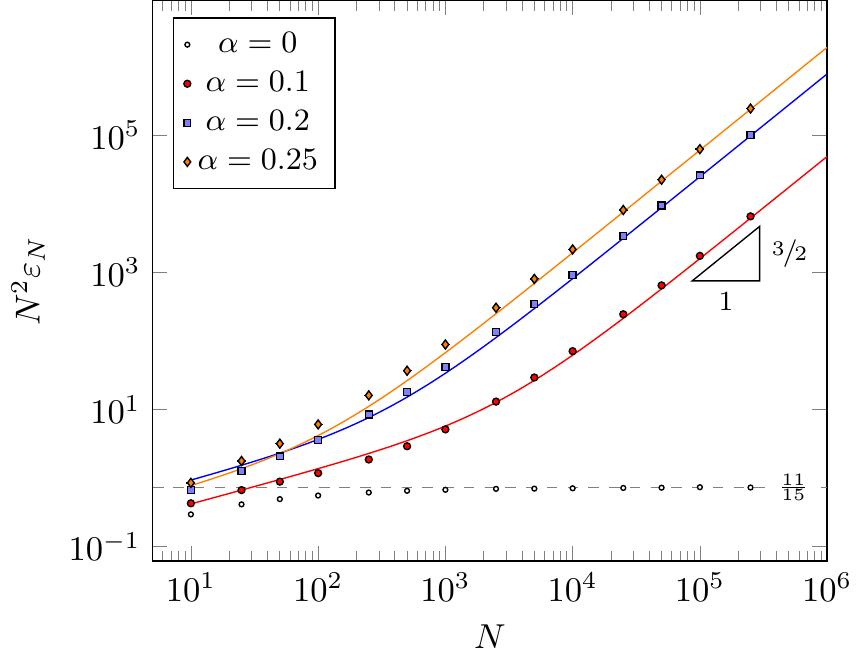}}
\caption{AOC in the case of of points distributed with the ``gapped'' PDF given in Eq.~\eqref{rhogap}. Error bars are represented but smaller than the markers. }
\end{figure}

\paragraph{Uniform distribution with a gap} To better exemplify this situation, let us consider the following PDF on $\Omega=[0,\sfrac{1}{2}-\sfrac{\alpha}{2}]\cup[\sfrac{1}{2}+\sfrac{\alpha}{2},1]$ with $\alpha\in[0,1)$ and $q\in(0,1)$,
\begin{equation}\label{rhogap}
 \varrho_{\alpha,q}(x)=
 \begin{cases}
  \frac{2q}{1-\alpha}&\text{if $x\in [0,\sfrac{1}{2}-\sfrac{\alpha}{2}]$,}\\
  \frac{2-2q}{1-\alpha}&\text{if $x\in [\sfrac{1}{2}+\sfrac{\alpha}{2},1]$},\\
  0&\text{otherwise}.
 \end{cases}
\end{equation}
A gap of width $\alpha$ is present in $\Omega=\bar\Omega$ when $\alpha\neq 0$. With reference to the notation adopted in this Section, in this case $\hat s=q$ for any value $\alpha\in(0,1)$ and therefore Eq.~\eqref{asygap} applies immediately, giving us
\begin{equation}\label{unigap}
 \varepsilon_N=2\alpha^p\sqrt{\frac{q(1-q)}{\pi N}}+o\left(\frac{1}{\sqrt{N}}\right).
\end{equation}
This scaling law has been numerically verified. In Fig.~\ref{fig:gap1} we consider $q=\sfrac{1}{2}$ and $p=2$, whereas in Fig.~\ref{fig:gap2} we assume $q=\sfrac{3}{4}$ and $p=4$, in both cases with different values of $\alpha$. The prediction in Eq.~\eqref{unigap} is in perfect agreement with the numerical results.

\section{Conclusions}\label{sec:3}
We have discussed the REAP on the line with convex weight cost $w(x,y)=|x-y|^p$ for $p>1$. We have assumed the points to be independently and randomly distributed on the line, according with a PDF $\varrho(x)\colon\mathds R\to\mathds R^+$. We have given a general expression for the asymptotic AOC $N^{\sfrac{p}{2}}\varepsilon_N$. We have shown that this general expression is possibly divergent, due to regions of very low density of points, i.e., to the zeros of $\varrho(x)$. We have given a recipe to take into account the effects of the discreteness of the problem when, denoting by $\Omega=\{x\in\mathds R\colon\varrho(x)>0\}$, the set $\bar\Omega\setminus\Omega$ is made up by isolated points (possibly including the point at infinity). We have then exemplified our approach, applying our recipe to a set of examples, extracting the exact scaling of the asymptotic AOC and, if possible, the coefficients appearing in it. Finally, we have considered the case in which the set $\Omega$ has gap, i.e., is composed by two disjoint intervals, showing that, in this situation, the effect of fluctuations in the number of points falling in each sub-interval dominates the asymptotic AOC.

The present work extends and completes the analysis given in Refs.~\cite{Caracciolo2015,Caracciolo2014c} for the quadratic REAP. The importance of the discussed regularization is however not restricted to the one-dimensional random assignment. Indeed, the same formulas for the AOC discussed here appear in other one-dimensional random optimization problems, such as the random Euclidean 2-matching~\cite{Caracciolo2018b} and the Traveling Salesman Problem~\cite{Caracciolo2018a} in the bipartite case. The understanding of the proper regularization to be adopted, when the simpler expression cannot be used and an anomalous scaling appears, is therefore relevant for a larger class of optimization problems, to which the analysis presented here can be applied. Finally, in Ref.~\cite{Caracciolo2015} an integral expression for the asymptotic AOC is given for $d> 1$. As the authors stress therein, the higher dimensional case might also require a regularization, depending on the properties of the domain and of the PDF adopted to extract the points. The criteria for such a regularization remain an open problem for future investigations.

\section*{Acknowledgments}
The authors are grateful to Luigi Ambrosio for useful discussions and correspondence, and for pointing us to Ref.~\cite{bobkov2014}. The authors also thank Giorgio Parisi, for many discussions on the assignment problem in Euclidean spaces. GS acknowledges the financial support of the Simons foundation (Grant No. 454949, Giorgio Parisi). The work presented in this article was supported by the project ``Meccanica statistica e complessit\`a'', a research grant funded by PRIN 2015 (Agreement no.~2015K7KK8L).

\section*{References}
\bibliographystyle{iopart-num}
\bibliography{biblio.bib}

\end{document}